\documentclass[9pt,twocolumn,twoside]{opticajnl}
\journal{opticajournal} 
\setboolean{shortarticle}{true}

\usepackage{lineno,bm}

\title{Polarization-preserving wavefront rotator}

\author[1,*]{Suman Karan}
\author[2]{Aman Srivastava}
\author[2]{Pratham Sachin Todkar}
\author[1,3]{Anand K. Jha}
\affil[1]{Department of Physics, Indian Institute of Technology Kanpur, Kanpur, UP 208016, India}
\affil[2]{Department of Computer Science and Engineering, Indian Institute of Technology Kanpur, Kanpur, UP 208016, India}

\affil[*]{sumankaran2@gmail.com}
\affil[3]{akjha@iitk.ac.in}

\begin{abstract}
A K-mirror rotates the wavefront of an incident optical field. However, the rotation always introduces polarization changes in the transmitted field. This is a serious concern for applications ranging from astronomical image derotation to orbital angular momentum spectrum characterization in photonic quantum technology.  Recent efforts have shown that the polarization change can be minimized significantly, but these require either a very small base angle that limits the field of view, or mirrors with a customized refractive index. Making the transmitted polarization state completely independent of the rotation angle has remained an open problem. In this work, we show that placing half-wave plates before and after a K-mirror and rotating them synchronously at half the K-mirror rotation angle makes the polarization change in the transmitted field exactly independent of the rotation angle. This works for any wavefront rotator, any base angle, any mirror refractive index, and any input state of polarization. We experimentally demonstrate the approach using a K-mirror with a base angle of $30^{\circ}$, which gives the largest field of view among practical designs, and find a mean polarization error of $\sim 1\%$, limited only by the retardance imperfection of commercially available half-wave plates. This has significant practical implications for applications that require precise wavefront rotation without polarization change.
\end{abstract}

\setboolean{displaycopyright}{false} 

\begin{document}

\maketitle

%
%
Rotating the wavefront of an optical field is always accompanied by polarization changes in the transmitted field\cite{moreno2004ao, karan2022ao}. This is a fundamental challenge in applications ranging from astronomy to high-dimensional orbital angular momentum (OAM) based quantum optics \cite{leach2004prl, pires2010prl}. In astronomical telescopes, wavefront rotators are used as image derotators in alt-azimuth mounts to counteract image rotation during sky tracking \cite{guo2014spie}. The rotation-dependent polarization change affects polarimetric and coronagraphic measurements \cite{schmid2018aa, holstein2023aa, anche2023aa}. In quantum optical experiments, wavefront rotation is central to OAM spectrum characterization via angular coherence measurement of optical fields, both at the high-light \cite{wang2017oe, gonzalez2006oe, pires2010ol, kulkarni2020prapp} and single-photon levels \cite{leach2004prl, leach2002prl, karan2025sciadv}. Any rotation-induced polarization change directly compromises the interferometric visibility and therefore the measurement fidelity \cite{pires2010prl, karan2025sciadv}. Apart from these applications, wavefront rotators find use in a broad range of other optical systems, including interferometry \cite{chu2008oe, mohanty1983optcomm}, microscopy \cite{zhi2015ol}, pattern recognition \cite{fujii1981optcomm}, and holography \cite{perezLopez2001ao}. Among the various devices used for wavefront rotation \cite{leach2004prl, padgett1999jmo, moreno2004ao}, K-mirrors are most suitable \cite{karan2022ao, peeters2007pra}. They consist of three mirrors with independent controls. This results in angular deviations of around 200 microradians due to rotation, which is significantly lower than the angular deviation of the order of milliradians for other commercially available wavefront rotators \cite{peeters2007pra}.

\begin{figure}[!t]
\centering
\includegraphics[scale=0.8]{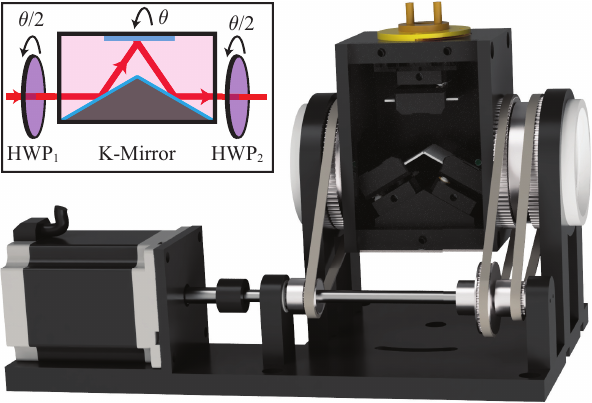}
\caption{Three-dimensional model of the home-built polarization-preserving wavefront rotator (PPWR) device for wavefront rotation. It consists of a K-mirror with two half-wave plates (HWP), ${\rm HWP}_1$ and ${\rm HWP}_2$, placed before and after the K-mirror, respectively. Both HWPs and the K-mirror are mounted on a common rotation stage driven by a stepper motor. The HWPs rotate at half the rotation angle of the K-mirror. The inset shows the conceptual schematic.}
\label{fig:conceptual_figure}
\end{figure}

Efforts have been made to reduce the polarization change induced by a rotating K-mirror. It was shown that the polarization change can be reduced from $30-50\%$ for a commercially available K-mirror \cite{karan2022ao} to as low as $3-10\%$ \cite{karan2024ao}. However, these approaches have important limitations. With a commercially available mirror coating, a near-zero polarization change requires a very small base angle of $17.88^{\circ}$. This significantly reduces the accessible field of view \cite{karan2024ao}. Moreover, at the optimal base angle of $30^{\circ}$, which gives the largest field of view, a mirror with a customized refractive index is needed. Such mirrors are not commercially available. In this context, achieving the transmitted state of polarization from a K-mirror, completely independent of the rotation angle for any base angle, any mirror refractive index, and any input state of polarization, has remained an open problem.

In this work, we show that placing half-wave plates (HWP) before and after a K-mirror and rotating them synchronously at half the K-mirror rotation angle makes the transmitted polarization state exactly independent of the rotation angle. We refer to this device as polarization-preserving wavefront rotator (PPWR). We present a Jones matrix analysis that analytically establishes this result. It is valid for any base angle, any mirror refractive index, and any input state of polarization. We further show that this HWP sandwich configuration, with the HWPs rotating at half of the rotator's angle, is universal and therefore works for any wavefront rotator, not just a K-mirror. We experimentally demonstrate this using a home-built K-mirror with a base angle of $30^{\circ}$. This gives the largest field of view among practical designs. We find a mean polarization error of $\sim 1\%$ for three different input states of polarization, namely linear, elliptical, and circular. Although theoretically the polarization error is expected to be exactly zero, experimentally it is limited only by the retardance imperfection of the commercially available half-wave plates.


Figure~\ref{fig:conceptual_figure} shows the design of our PPWR device, which consists of a K-mirror with two half-wave plates, ${\rm HWP}_1$ and ${\rm HWP}_2$, placed before and after it, respectively. All three components are mounted on a common rotation stage driven by a stepper motor, such that the HWPs rotate synchronously by $\theta/2$ when the K-mirror is rotated by $\theta$. In order to analyze the transmitted states of polarization, we use the Jones matrix formalism. The incident electric field is written as ${\bm E}^{\rm in} = E^{\rm in}_x~ \hat{\bm x} + E^{\rm in}_y~\hat{\bm y}$, where  $E^{\rm in}_x = \cos \psi_{\rm in}$ and $E^{\rm in}_y = \sin \psi_{\rm in}e^{i \delta_{\rm in}}$. The Jones matrix of the K-mirror at rotation angle $\theta$ as given in Ref.~\cite{karan2022ao}, is
\begin{equation}\label{eqn:KM_matrix}
T_{\rm KM}\left(\theta \right) = \begin{bmatrix} T^{s}_{\rm KM}\cos^2 \theta +  T^{p}_{\rm KM}\sin^2 \theta & \left(T^{s}_{\rm KM}-T^{p}_{\rm KM}\right)\sin\theta \cos \theta \\
\left(T^{s}_{\rm KM}-T^{p}_{\rm KM}\right)\sin\theta \cos \theta &  T^{p}_{\rm KM}\cos^2 \theta +  T^{s}_{\rm KM}\sin^2 \theta \end{bmatrix},
\end{equation}
\begin{figure}[!t]
\centering
\includegraphics[scale=0.88]{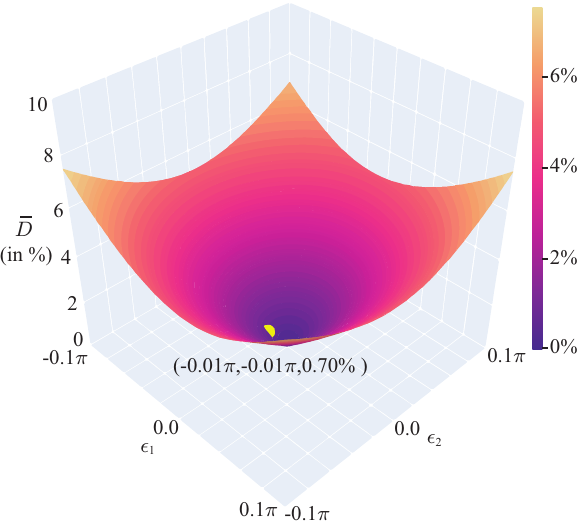}
\caption{Numerically simulated plot of mean polarization errors $\bar{D}$ in $\%$ as a function of the retardance errors $\epsilon_1$ and $\epsilon_2$ of HWP1 and HWP2, respectively, for a linearly polarized input state. The yellow dot marks the operating point corresponding to commercially available Thorlabs HWPs.  
}
\label{fig:hwp_simulation}
\end{figure}
where $\theta= 0^{\circ}$ is defined when  all three mirrors of the K-mirror are in  the ${\bm xz}$ plane. The forms of  $T_{\rm KM }^{s} $ and $T_{\rm KM }^{p}$ are given in Refs.~\cite{karan2024ao, karan2022ao} and depend on the base angle $\beta$ of the K-mirror and the refractive index $n_M$ of the mirror coating. For a half-wave plate with retardance $\pi + \epsilon$, where $\epsilon$ is the  retardance error, the Jones matrix with the fast axis at angle $\theta/2$ with respect to $\hat{\bm{x}}$ can be expressed as 
\begin{equation}\label{eqn:hwp}
{\rm HWP}^{\epsilon}\left(\theta\right) = \begin{bmatrix} \cos^2 \theta - e^{i\epsilon}\sin^2 \theta  &  \left(1 + e^{i \epsilon}\right)\sin \theta \cos \theta \\
\left(1 + e^{i \epsilon}\right)\sin \theta \cos \theta \ &  \sin^2 \theta - e^{i\epsilon}\cos^2 \theta   \end{bmatrix}.
\end{equation}
If the retardance errors of ${\rm HWP_1}$ and ${\rm HWP_2}$ are $\epsilon_1$ and $\epsilon_2$, respectively, the Jones matrix of the PPWR at rotation angle $\theta$ can be written as
\begin{equation}\label{eqn: trasfer ppwr}
T_{\rm PPWR}\left(\theta \right) ={\rm HWP}^{\epsilon_2}_{2}\left(\theta/2 \right) \times T_{\rm KM}\left(\theta \right)\times {\rm HWP}^{\epsilon_1}_{1}\left(\theta/2 \right).
\end{equation}
The transmitted field $ {\bm E}^{\rm out} = E^{\rm out}_x~ \hat{\bm x} + E^{\rm out}_y~\hat{\bm y}$ at any rotation angle $\theta$ of the PPWR is given by $ {\bm E}^{\rm out} = T_{\rm PPWR}\left(\theta \right) {\bm E}^{\rm in}$.

For ideal half-wave plates, $\epsilon_1 = \epsilon_2 = 0$, and Eq.~(\ref{eqn: trasfer ppwr}) reduces to
\begin{equation} \label{eqn:ideal_transfer_ppwr}
T_{\rm PPWR}\left(\theta \right) = \begin{bmatrix} T^{s}_{\rm KM} & 0 \\
0 & T^{p}_{\rm KM } \end{bmatrix},
\end{equation}
where $T_{\rm PPWR}\left(\theta \right)$  is independent of $\theta$, and thus the transmitted state of polarization is exactly independent of the rotation angle. This holds for any base angle, any mirror refractive index, and any input state of polarization. From Eq.~(\ref{eqn:ideal_transfer_ppwr}), we infer that for a horizontal (H) linearly polarized input state $\left[ 1, 0 \right]^{\rm T}$, the transmitted states for any arbitrary $\theta$ are $\left[ T^{s}_{\rm KM}  , 0\right]^{\rm T}$, and for a vertical (V) linearly polarized input state $\left[ 0, 1 \right]^{\rm T}$, the transmitted states are $\left[0, T^{p}_{\rm KM} \right]^{\rm T}$. Since $T^{s}_{\rm KM}$ and $T^{p}_{\rm KM}$ are just two complex numbers, they have no effect on the polarization state, and thereby the transmitted states are the same as the respective input states. We further note that for input states of polarization other than these two special cases, the transmitted states of polarization will not be the same as the incident state.

\begin{figure}[!b]
\centering
\includegraphics[scale=0.9]{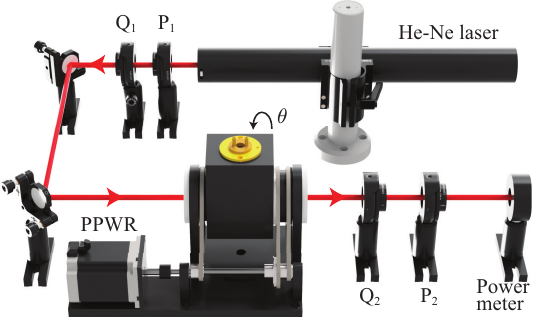}
\caption{Schematic of the experimental setup to measure the transmitted state of polarization at different rotation angles of the polarization-preserving wavefront rotator (PPWR) device. ${P_1}$, ${ P_2}$: polarizers; $Q_1$, $Q_2$: quarter-wave plates.  }
\label{fig:exp_setup}
\end{figure}
\begin{figure*}[!t]
\centering
\includegraphics[scale=1]{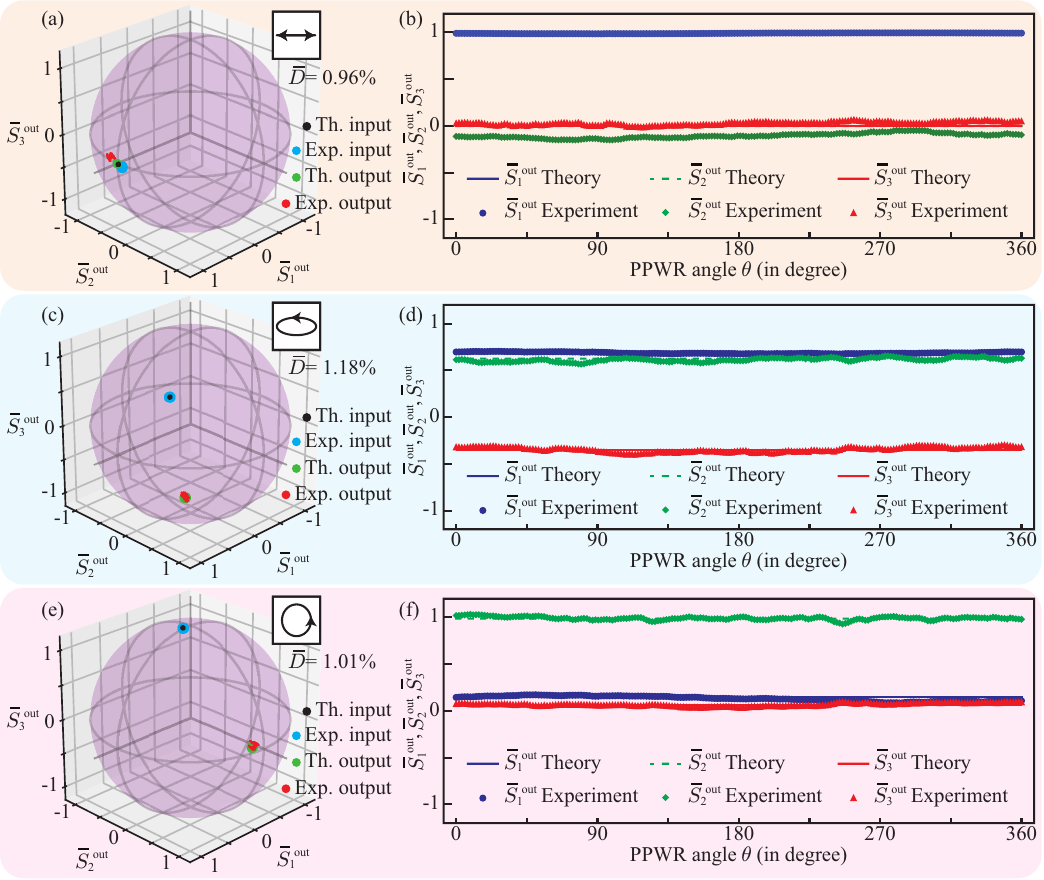}
\caption{Poincar\'e sphere representation of the transmitted states of polarization at different rotation angles of the PPWR device for three different input states of polarization: (a) linear, (c) elliptical, and (e) circular. The corresponding normalized Stokes parameters are shown in (b), (d), and (f), respectively. The incident state of polarization in each case is shown in the inset.  }
\label{fig:exp_result}
\end{figure*}

The state of polarization of  ${\bm E}^{{\rm in}(\rm out) }$ can be uniquely represented by four Stokes parameters: $S_0^{{\rm in}(\rm out) } = |E_x^{{\rm in}(\rm out) }|^2 + |E_y^{{\rm in}(\rm out) }|^2$,  $S_1^{{\rm in}(\rm out) } = |E_x^{{\rm in}(\rm out) }|^2 - |E_y^{{\rm in}(\rm out) }|^2$, $S_2^{{\rm in}(\rm out) } = 2 {\rm Re}\left[E_x^{{\rm in}(\rm out) *} E_y^{{\rm in}(\rm out)}\right]$,  and $S_3^{{\rm in}(\rm out) } = 2 {\rm Im}\left[E_x^{{\rm in}(\rm out) *} E_y^{{\rm in}(\rm out)}\right]$ \cite{goldstein2017crc}. Here ${}^*$ denotes complex conjugate. 
For a state with degree of polarization equal to $1$, the normalized Stokes parameters are: ${\bar S}_1^{{\rm in}(\rm out)} ={S}_1^{{\rm in}(\rm out)}/{S}_0^{{\rm in}(\rm out)} $,  ${\bar S}_2^{{\rm in}(\rm out)} ={S}_2^{{\rm in}(\rm out)}/{S}_0^{{\rm in}(\rm out)} $, and  ${\bar S}_3^{{\rm in}(\rm out)} ={S}_3^{{\rm in}(\rm out)}/{S}_0^{{\rm in}(\rm out)} $. These normalized Stokes parameters $\left({\bar S}_1^{{\rm in}(\rm out)}, {\bar S}_2^{{\rm in}(\rm out)}, {\bar S}_3^{{\rm in}(\rm out)} \right)$ can be represented as a point on the surface of the Poincar\'e sphere, where each point uniquely represents a particular state of polarization.

Ideally, if the transmitted state is completely rotation-independent, all states measured at rotation angles from $0^{\circ}$ to $360^{\circ}$ map to a single point on the surface of the Poincar\'e sphere. Otherwise, they form a loop, as shown in Ref.~\cite{karan2022ao}. The smaller the loop, the smaller the polarization change due to rotation. As Eq.~(\ref{eqn:ideal_transfer_ppwr}) shows, the Jones matrix of the PPWR has no $\theta$ dependence, and therefore the transmitted states of polarization map to a single point on the Poincar\'e sphere.

Inspired by the metric defined in Ref.~\cite{karan2022ao}, we define the mean polarization error $\bar{D}$ as the mean geodesic distance on the surface of the Poincar\'e sphere between the transmitted state of polarization at $\theta = 0^{\circ}$ and the transmitted states of polarization at all other rotation angles $\theta$ of PPWR. We express $\bar{D}$ in percent form as  
\begin{equation} \label{eqn:Dbar}
\bar{D} =\frac{\int_{\theta=0}^{2\pi} \cos^{-1} \left[ \sum_{i=1}^{3} \bar{S}^{\rm{out}}_{i}\,\bar{S}^{\rm{out}0}_{i} \right] \sqrt{ \sum_{i=1}^{3} \left( \frac{d S^{\rm{out}}_{i}}{d\theta} \right)^2} d\theta}{ \int_{\theta=0}^{2\pi} \sqrt{ \sum_{i=1}^{3} \left( \frac{d S^{\rm{out}}_{i}}{d\theta} \right)^2} d\theta }\%,
\end{equation}
where $\bar{S}^{\rm{out}0}_{i}$, with  $i \in \left\lbrace  1,~2,~3\right \rbrace$ are the normalized Stokes parameters of the transmitted state of polarization at $\theta=0^{\circ}$. When the transmitted states of polarization are completely rotation-independent, they map to a single point on the Poincar\'e sphere. In this case, Eq.~(\ref{eqn:Dbar}) takes the $0/0$ form, and from continuity we define $\bar{D}$ as $0\%$. Here, $\bar{D} = 0 \%$ represents transmitted states of polarization completely independent of rotation, and $\bar{D} = 100 \%$ corresponds to the maximum possible polarization change due to rotation.

Figure~\ref{fig:hwp_simulation} shows the plot of numerically simulated $\bar{D}$ as a function of the retardance errors $\epsilon_1$ and $\epsilon_2$ of HWP1 and HWP2, respectively. Here, both $\epsilon_1$ and $\epsilon_2$ range from $-0.1\pi$  to $0.1\pi$. This shows $\bar{D}=0\%$ for $\epsilon_1= \epsilon_2=0$, and $\bar{D}$ increases as the retardance errors increase. For the retardance error of $\epsilon_1 =\epsilon_2 =-0.01\pi$ corresponding to our Thorlabs HWPs (WPHSM05-633), the simulation predicts $\bar{D}$  to be $0.70\%$. The yellow dot in Fig.~\ref{fig:hwp_simulation} represents this point. We further show that this configuration is universal and works for any wavefront rotator, not just a K-mirror, with $\alpha = 1/2$ being the unique value for polarization preservation (see Supplement~1).


Figure~\ref{fig:exp_setup} shows the experimental setup to  measure the Stokes parameters of the transmitted states of polarization at different rotation angles $\theta$ of the PPWR and thereby $\bar{D}$. A 5 mW Newport He-Ne laser of wavelength 632.8 nm is sent through a linear polarizer $P_1$ to generate H-polarized light. We then insert a quarter-wave plate $Q_1$ with its fast axis rotated at $24^{\circ}$ and $45^{\circ}$ to generate elliptically and circularly polarized input states, respectively. In assembling our home-built PPWR, we use Thorlabs 1" protected silver mirrors PF10-03-P01 and zero-order half-wave plates WPHSM05-633 as ${\rm HWP}_1$ and ${\rm HWP}_2$. Using standard Stokes parameter measurement via polarizer $P_2$, quarter-wave plate $Q_2$ and a power meter, we measure the transmitted states of polarization at different rotation angles $\theta$ of the PPWR.

Figure~\ref{fig:exp_result} presents the experimental results. The Poincar\'e sphere representation of the experimentally measured transmitted states of polarization for rotation angles $\theta$ from $0^{\circ}$ to $360^{\circ}$ in steps of $1.8^{\circ}$ for a linearly polarized incident state of polarization is shown in Fig.~\ref{fig:exp_result}(a). The theoretical incident polarization state is denoted by a black dot on the Poincar\'e sphere. The experimentally measured incident state of  polarization is shown by the blue dot and in the inset. The theoretically predicted transmitted states of polarization are shown by the green loop and the corresponding experimentally measured states of polarization are shown by the red loop. Figure~\ref{fig:exp_result}(b) shows the theoretically expected and experimentally measured normalized Stokes parameters: $\bar{S}^{\rm out}_1$, $\bar{S}^{\rm out}_2$, and $\bar{S}^{\rm out}_3$ as a function of $\theta$ for the same linearly polarized input state. The corresponding plots for the elliptically and circularly polarized incident states of polarization are shown in Figs.~\ref{fig:exp_result}(c),~\ref{fig:exp_result}(d) and Figs.~\ref{fig:exp_result}(e),~\ref{fig:exp_result}(f), respectively. The measured mean polarization errors for linearly, elliptically, and circularly polarized input states are $0.96\%$, $1.18\%$, and $1.01\%$, respectively, all consistent with the theoretically predicted value of $0.70\%$.

In conclusion, we have proposed and demonstrated a wavefront rotator, the PPWR, that rotates the wavefront while keeping the transmitted state of polarization completely independent of the rotation angle. The PPWR  consists of a K-mirror placed between two HWPs, assembled such that when the K-mirror is rotated by  $\theta$, the HWPs rotate by $\theta/2$. We  have shown that this result holds for any mirror refractive index, any base angle, and any input state of polarization. We have demonstrated that for protected silver mirror coating and a base angle of $30^{\circ}$, which gives the largest field of view, the mean polarization error is only $\sim 1\%$ for linearly, elliptically, and circularly polarized input states.
This is due to the retardance imperfection of the commercially available HWPs. Otherwise, the mean polarization error is exactly zero. We have also shown that for H- and V-polarized input states, the transmitted states of polarization not only remain the same for all rotation angles but also coincide with the input state of polarization. This is achieved without the need for any additional optical elements, which is crucial for interferometry-based applications \cite{karan2025sciadv, pires2010prl}. Furthermore, we have shown that the HWP sandwich configuration is universal and works for any wavefront rotator, not just a K-mirror. We have further shown that $\alpha = 1/2$, that is, rotating the HWPs at half the rotator's angle, is the unique value for which this geometry becomes polarization-preserving. Our work has significant implications for image derotation and orbital angular momentum based high-dimensional quantum information applications where wavefront rotation without polarization change is essential.

\begin{backmatter}
\bmsection{Funding} We acknowledge financial support from the Science and Engineering Research Board
through grants STR/2021/000035 \& CRG/2022/003070, and from the Department of Science \& Technology, Government of India through grant DST/ICPS/QuST/Theme-1/2019 and through the National Quantum Mission (NQM) technical group project on quantum imaging.

\bmsection{Disclosures} The authors declare no conflicts of interest.

\bmsection{Data availability} Data underlying the results presented in this paper are not publicly available at this time but may be obtained from the authors upon reasonable request.

\bmsection{Supplemental document}
See Supplement 1 for supporting content.

\end{backmatter}

\bibliography{ir_zero_pol_ref}

\bibliographyfullrefs{ir_zero_pol_ref}

\pagebreak
\clearpage

\appendix
\onecolumn
\setcounter{equation}{0}
\setcounter{figure}{0}
\setcounter{table}{0}
\renewcommand{\theequation}{S\arabic{equation}}
\renewcommand{\thefigure}{S\arabic{figure}}
\renewcommand{\thetable}{S\arabic{table}}
\renewcommand{\thesection}{S\arabic{section}}

\begin{center}
{\LARGE \textbf{Polarization-preserving wavefront rotator: supplemental document}}

\vspace{0.5em}
Suman Karan et al.
\vspace{1em}
\end{center}

\section{ Universality of the polarization-preserving geometry}
The geometry of placing HWPs before and after a wavefront rotator and rotating them at half the rotator's angle is not specific to K-mirrors. In this section, we  show that the proposed and demonstrated configuration is universal and works for any wavefront rotator. The Jones matrix of the K-mirror in Eq.~(1) of the main manuscript can be written as
\begin{equation}\label{eqn:general_form}
T_{\rm KM}\left(\theta \right) = R\left(\theta \right) T_{\rm KM}\left(0 \right) 
R^{T}\left(\theta \right),
\end{equation}
where $R\left(\theta \right) = \begin{bmatrix} \cos\theta & -\sin\theta \\ 
\sin\theta & \cos\theta \end{bmatrix}$ is the rotation matrix and 
$T_{\rm KM}\left(0\right)$ is the Jones matrix of the K-mirror at $\theta = 0$. 
This form holds for any wavefront rotator, including Dove prisms \cite{moreno2004ao, karan2022ao}, Delta prism \cite{moreno2004ao}, Pechan prism \cite{moreno2004ao}, 
and five-mirror image derotators \cite{hou2018oe}, differing only in their Jones matrix elements  at $\theta = 0$. For the configuration in which two HWPs are placed before and after a wavefront rotator  and rotated by $\alpha\theta$ when the rotator is rotated by $\theta$, the Jones matrix of the full configuration is
\begin{equation}\label{eq:tot_config}
T_{\rm tot}\left(\alpha, \theta\right) = {\rm HWP}\left(\alpha\theta \right) 
R\left(\theta \right) T\left(0 \right) R^{T}\left(\theta \right) 
{\rm HWP}\left(\alpha\theta \right),
\end{equation}
where ${\rm HWP}\left(\alpha\theta \right) = \begin{bmatrix} \cos 2\alpha\theta & 
\sin 2\alpha\theta \\ \sin 2\alpha\theta & -\cos 2\alpha\theta \end{bmatrix}$ 
is obtained from Eq.~(2) of the main manuscript with $\epsilon = 0$. We show that
\begin{equation}\label{eqn:hwp_rotation_identity}
{\rm HWP}\left(\alpha\theta \right) R\left(\theta \right) =R^{T}\left(\theta \right){\rm HWP}\left(\alpha\theta \right) =
{\rm HWP}\left(\frac{2\alpha-1}{2}\theta\right).
\end{equation}
Substituting into Eq.~(\ref{eq:tot_config}), we get
\begin{equation}\label{eqn:tot_simplified}
T_{\rm tot}\left(\alpha, \theta\right) = {\rm HWP}\left(\frac{2\alpha-1}{2}
\theta\right) T\left(0\right) {\rm HWP}\left(\frac{2\alpha-1}{2}\theta\right).
\end{equation}
For $\alpha = 1/2$, this reduces to
\begin{equation}\label{eqn:universal_result}
T_{\rm tot}\left(1/2, \theta\right) = {\rm HWP}\left(0\right) T\left(0\right) 
{\rm HWP}\left(0\right),
\end{equation}
which is independent of $\theta$. This holds for any $T\left(0\right)$, and 
therefore for any wavefront rotator. We note that $\alpha = 1/2$ is the unique 
value for which $T_{\rm tot}$ becomes independent of $\theta$. For this value, and $T\left(0\right)$ of the K-mirror, $T_{\rm tot}\left(1/2, \theta\right) = T_{\rm PPWR}\left(\theta\right) $, consistent with Eq.~(4) of the main manuscript. This confirms that the polarization-preserving wavefront rotator (PPWR) is a specific implementation of this universal polarization-preserving geometry.


\end{document}